\documentclass[aip,jcp,reprint,amsmath,amssymb]{revtex4-2}
\newcommand{\vv}[1]{\mathbf{#1}}
\renewcommand{\d}[1]{\ensuremath{\operatorname{d}\!{#1}}}
\usepackage{graphicx}
\usepackage[suffix=]{epstopdf}
\usepackage{natmove}
\usepackage{xcolor}

\begin{document}
\title{relentless: Transparent, reproducible molecular dynamics simulations for optimization}

\author{Adithya N Sreenivasan}
\thanks{These authors contributed equally.}
\affiliation{McKetta Department of Chemical Engineering, The University of Texas at Austin, Austin, TX 78712, USA}
\affiliation{Department of Chemical Engineering, Massachusetts Institute of Technology, Cambridge, MA 02139, USA}

\author{C. Levi Petix}
\thanks{These authors contributed equally.}
\affiliation{Department of Chemical Engineering, Auburn University, Auburn, AL 36849, USA}

\author{Zachary M. Sherman}
\affiliation{Department of Chemical Engineering, University of Washington, Seattle, WA 98195, USA}

\author{Michael P. Howard}
\email{mphoward@auburn.edu}
\affiliation{Department of Chemical Engineering, Auburn University, Auburn, AL 36849, USA}

\begin{abstract}
relentless is an open-source Python package that enables the optimization of objective functions computed using molecular dynamics simulations. It has a high-level, extensible interface for model parametrization; setting up, running, and analyzing simulations natively in established software packages; and gradient-based optimization. We describe the design and implementation of relentless in the context of relative entropy minimization, and we demonstrate its abilities to design pairwise interactions between particles that form targeted structures. relentless aims to streamline the development of computational materials design methodologies and promote the transparency and reproducibility of complex workflows integrating molecular dynamics simulations.
\end{abstract}

\maketitle

\section{Introduction}
Classical molecular dynamics (MD) is a popular computational tool for investigating a variety of materials \cite{allen:2017, rapaport:2004}. For example, MD simulations are routinely performed to study the thermophysical properties of pure substances and mixtures, the rheology of nanoparticle suspensions and polymer solutions\cite{chen:macro:2018, nikoubashman:macro:2017}, and the behavior of peptides\cite{ulmschneider:acr:2018} and proteins\cite{karplus:pnas:2005}. The widespread availability of open-source software enabling different stages of a typical MD workflow has greatly simplified the process of performing molecular simulations. For example, AmberTools\cite{case:jcheminfmod:2023}, foyer\cite{klein:compmatsci:2019} \& mbuild\cite{klein:2016}, PACKMOL\cite{martinez:jcc:2009}, and QwikMD\cite{ribeiro:scirep:2016} can create initial configurations and other simulation inputs. Packages such as GROMACS\cite{abraham:softwareX:2015}, HOOMD-blue\cite{anderson:compmatsci:2020}, LAMMPS\cite{thompson:compphyscom:2022}, NAMD\cite{phillips:jcp:2020}, and OpenMM\cite{eastman:jctc:2013} can perform the simulations and have been optimized for high-performance computing resources. Visualization tools such as OVITO\cite{stukowski:modsimmatscieng:2009} and VMD\cite{humphrey:jmolgraph:1996} can render simulation outputs, while analysis packages such as freud\cite{ramasubramani:comphyschem:2020}, mdanalysis\cite{michaud:jcompchem:2011, gowers:propysci:2016}, and mdtraj\cite{mcgibbon:jbiophys:2015} can calculate measurable properties from simulation trajectories. In addition to their convenience, these open-source, community-supported tools increase transparency and reproducibility of MD simulation workflows compared to using private codes.

Beyond direct application as computational experiments, there is increasing interest in using MD simulations as components of larger optimization workflows\cite{sherman:jcp:2020, dePablo:compmat:2019, kadulkar:arcbe:2022}. An MD simulation can be considered a function whose measurable outputs can be optimized with respect to its inputs. Examples of such use cases are fitting of force field parameters to experimentally measured properties\cite{befort:jcim:2021, wang:jctc:2023}, bottom-up coarse-graining\cite{noid:jcp:2008, shell:jcp:2008,shell:2016, chaimovich:jcp:2011}, and computational materials design\cite{sherman:jcp:2020,long:molsystdeseng:2018, miskin:pnas:2016, torquato:softmat:2009,jain:aiche:2014, jadrich:jcp:2017, ferguson:jpcm:2017, gadelrab:msde:2017,jackson:coice:2019, dijkstra:natmat:2021}. The software we have developed is able to optimize different functions computed from MD simulations, but in this paper, we focus on using simulations to design pairwise interactions between particles in order to produce a specified (target) structure. This type of inverse design for self-assembly has received considerable interest both as a fundamental scientific problem and for its potential to discover novel, tailored materials\cite{torquato:softmat:2009,jain:softmatter:2013, jain:aiche:2014, jain:physrev:2014, engel:natmat:2015,jadrich:softmatter:2015,lindquist:jcp:2016, jadrich:jcp:2017,lindquist:jpcb:2018, pineros:jcp:2018,lindquist:softmatter:2017, sherman:jcp:2020, gadelrab:msde:2017, ferguson:jpcm:2017, jackson:coice:2019,  banerjee:jcp:2019,ma:softmatter:2019, dijkstra:natmat:2021}.

We will carry out our design process by minimizing the relative entropy \cite{shell:jcp:2008,shell:2016} from a statistical mechanical ensemble associated with a target structure to the ensemble that is simulated using the current best-guess for the particle interactions \cite{lindquist:jcp:2016, jadrich:jcp:2017}. Mathematical details are discussed in Section \ref{sec:REM}, but a typical workflow is: (1) set up an MD simulation using the current particle interactions, (2) perform the simulation, (3) compute the gradient of the relative entropy from the simulation trajectory, and (4) update the particle interactions based on this gradient until convergence. For best performance, the simulations in Step 2 should usually be done in an established MD package, while the other steps of the workflow can be implemented in a wrapper code that does the other stages of analysis and drives the optimization. For example, VOTCA\cite{ruhle:jctc:2009, ruhle:macrotheosim:2011, mashayak:plosone:2015, scherer:jpccp:2018} implements relative entropy minimization for coarse-graining and can perform Step 2 natively in GROMACS or LAMMPS, but implements the other steps using its own analysis routines, file-based inputs, and scripting interface. As another example, OpenMSCG implements relative entropy minimization through a combined Python and command-line interface, with the user supplying an MD simulation command to be run in a particular package as a subprocess\cite{peng:jpcb:2023}.

\begin{figure*}
    \includegraphics{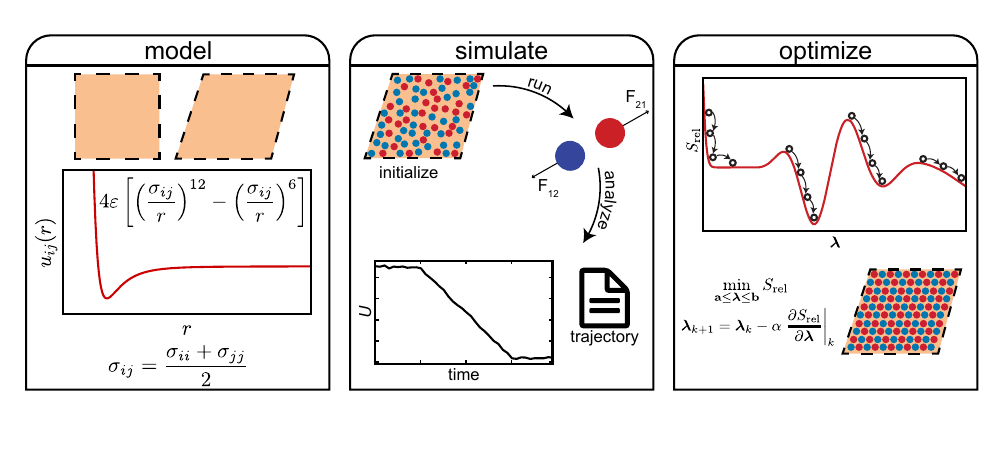}
    \caption{relentless is organized into three modules. \textbf{model} defines the thermodynamic ensemble (e.g., temperature, number of particles, simulation box), potential energy functions, and model parameters and their relationships (e.g., mixing rules). \textbf{simulate} initializes, runs, and analyzes simulations. \textbf{optimize} minimizes an objective function computed from a simulation.}
    \label{fig:overview}
\end{figure*}
Many simulation packages now provide their own Python interfaces, which in principle allows the entire relative entropy minimization workflow to be carried out by connecting various tools that already exist in the large scientific Python ecosystem. However, these simulation interfaces and the simulation outputs do not yet have an accepted standard, so the workflow would need to be tailored to the simulation package being used, limiting reusability if another package (or sometimes even version of the same package!) is preferred. OpenMM defines a hardware-independent interface for performing simulations that is highly extensible by users \cite{eastman:jctc:2013}; however, its high-level interface is limited to the hardware platforms that are implemented, and users may have a preference to a different simulation package for a variety of reasons such as the hardware or software they have access to. The Atomic Simulation Environment (ASE)\cite{larsen:condmat:2017} can perform MD simulations through a common Python interface that calculates the forces on particles using different simulation packages, but the full performance of these packages may be underutilized because ASE time-integrates the particle coordinates internally. The Python Simulation Interface for Molecular Modeling (pysimm)\cite{fortunato:softwarex:2017} provides a Python wrapper for setting up and running simulations, primarily of amorphous polymers, natively in LAMMPS, but it does not currently support other popular MD simulation packages and its interface would be challenging to generalize to do so.

Drawing inspiration from these and other prior and ongoing efforts, we have developed relentless as a transparent, reproducible, usable by others, and extensible (TRUE)\cite{thompson:jmolphys:2020} Python package enabling optimization of objective functions computed using MD simulations. Toward this aim, we have developed a high-level interface for setting up, running, and analyzing MD simulations that is simple but extensible to meet users' specific needs. relentless is organized around three concepts of typical workflows: model, simulate, and optimize (Fig.~\ref{fig:overview}). The model specifies the simulation ensemble and molecular interactions, including variables used to represent model parameters. Variables that will be optimized can be box-constrained based on physical knowledge (e.g., to be positive), and models can be parametrized on combinations of variables (e.g., arithmetic mean) when such relationships are known. The simulation protocol is prescribed through common initialization, simulation, and analysis operations. These operations are translated to a specific simulation package when the simulation is run; relentless currently supports multiple versions of HOOMD-blue and LAMMPS. An optimization problem is defined in relentless by an objective function that can run simulations if needed, and relentless can minimize this objective function using gradient-based optimization with user-defined stopping criteria. Core concepts in relentless such as the molecular interactions, simulation operations and packages, and objective functions are designed to be easily extended to new applications using best practices of object-oriented programming.

The rest of this article is organized as follows. We first summarize some mathematical background on relative entropy minimization, with an emphasis on the points that have informed our software design (Section~\ref{sec:REM}). We then give an overview of the elements of relentless that we believe help achieve a TRUE approach to performing MD simulations for optimization (Section~\ref{sec:design}). We demonstrate relentless on two example problems: (1) fitting the diameters of a binary hard-sphere fluid and (2) finding interactions that self-assemble cubic diamond (Section~\ref{sec:examples}). We conclude with discussion in Section \ref{sec:Discussion}.

\section{Relative-entropy minimization \label{sec:REM}}
The relative entropy $S_{\rm rel}$ was developed as a technique for fitting coarse-grained models to fine-resolution simulation data \cite{shell:2016, shell:jcp:2008}, then was later adapted to the analogous inverse design problem of finding particle interactions that produce a certain structure \cite{lindquist:jcp:2016, jadrich:jcp:2017}. The relative entropy for inverse design in the canonical ensemble (constant number of particles $N$, volume $V$, and temperature $T$) is 
\begin{equation}
    S_{\rm{rel}}(\boldsymbol{\lambda}) = \int \d{\vv{r}^N} p_0({\vv{r}^N}) \ln\left[\frac{p_0({\vv{r}^N})}{p({\vv{r}^N;\boldsymbol{\lambda}})} \right], 
    \label{eq:srel}
\end{equation}
where $\vv{r}^N = \{ \vv{r}_1, \dots, \vv{r}_N \}$ is the configuration of the $N$ particles, $\vv{r}_i$ is the position of particle $i$, $p_0(\vv{r}^N)$ is the desired probability density function for observing a configuration (the target ensemble), and $p(\vv{r}^N;\boldsymbol{\lambda})$ is the probability density function for observing the same configuration with particle interactions parametrized by $\boldsymbol{\lambda}$ (the simulation ensemble). For inverse design, the target ensemble is usually produced artificially, e.g., by tethering point particles to sites of a crystalline lattice with springs and creating an ensemble of structures at a nominal temperature. The parameters that make up $\boldsymbol{\lambda}$ can be physical (e.g., particle diameter or interaction energy \cite{lindquist:softmatter:2016, lindquist:jcp:2021}) or artificial (e.g., knots of a spline potential \cite{pineros:jcp:2018, lindquist:jcp:2016, lindquist:jpcb:2018, pretti:jcp:2021}). In both cases, the values of $\boldsymbol{\lambda}$ may be constrained to certain ranges, e.g., to enforce experimentally known bounds or to force the interactions to have certain properties\cite{lindquist:jcp:2016, lindquist:jpcb:2018}. Some parameters may also be related to others, for example, by applying mixing rules to obtain cross-interaction parameters from self-interaction parameters\cite{allen:2017}.

The relative entropy is nonnegative and it is zero when $p = p_0$, so particle interactions that produce a simulation ensemble similar to the target ensemble can be designed by minimizing $S_{\rm{rel}}$ with respect to $\boldsymbol{\lambda}$. For the canonical ensemble, it can be shown that at thermodynamic equilibrium
\begin{equation}
    S_{\rm{rel}} = \beta \left\langle U(\vv{r}^N; \boldsymbol{\lambda}) - U_0(\vv{r}^N) \right\rangle_0 + \ln Z(\boldsymbol{\lambda}) - \ln Z_0,
    \label{eq:srelavg}
\end{equation}
where $U$ is the total potential energy of a configuration in the simulated ensemble and
\begin{equation}
    Z(\boldsymbol{\lambda}) = \int \d{\vv{r}^N} \exp[-\beta U(\vv{r}^N;\boldsymbol{\lambda})]
\end{equation}
is the canonical partition function for the simulated ensemble; both $U$ and $Z$ depend on $\boldsymbol{\lambda}$. $U_0$ and $Z_0$ are analogous quantities for the target ensemble that do not depend on $\boldsymbol{\lambda}$, and $\langle \dots \rangle_0$ denotes an average in the target ensemble. Evaluation of $S_{\rm rel}$ itself with simulations is difficult because of the partition functions in eq.~\eqref{eq:srelavg}, but the gradient of $S_{\rm{rel}}$ is more easily obtained. Differentiation of eq.~\eqref{eq:srelavg} with respect to $\boldsymbol{\lambda}$ eliminates $Z_0$ and $U_0$, which are not functions of $\boldsymbol{\lambda}$, and leaves
\begin{equation}
\frac{\partial S_{\rm rel}}{\partial \boldsymbol{\lambda}} =  \beta \left( \left\langle \frac{\partial U}{\partial \boldsymbol{\lambda}} \right\rangle_0 - \left\langle \frac{\partial  U}{\partial \boldsymbol{\lambda}} \right\rangle \right),
\label{eq:gradgen}
\end{equation}
where $\langle \dots \rangle$ denotes an average in the simulated ensemble. Hence, eq.~\eqref{eq:gradgen} can be evaluated by performing a simulation with a given $\boldsymbol{\lambda}$ and averaging $\partial U/\partial \boldsymbol{\lambda}$.

Equation~\ref{eq:srelavg} is valid for any potential energy function, but for many models, the potential energy $U$ can be considered to be a sum of isotropic pairwise interactions $u^{(ij)}(r)$ between two particles of types $i$ and $j$ separated by distance $r$. For a system with $n$ types,
\begin{equation}
    U(\vv{r}^N; \boldsymbol{\lambda}) = \frac{1}{2} \sum_{i=1}^n \sum_{j = 1}^n \sum_{k=1}^{N_i} \sum_{l = 1}^{N_j} (1-\delta_{ij}\delta_{k l})u^{(ij)}(r_{kl}; \boldsymbol{\lambda}),
    \label{eq:U}
\end{equation}
where $N_i$ is the number of particles of type $i$ ($N = \sum_{i=1}^n N_i$), $r_{kl}$ is the distance between particles $k$ and $l$ of types $i$ and $j$, and $\delta$ is the Kronecker delta. We will discuss considerations for defining pair potentials $u^{(ij)}$ in Section \ref{sec:design:model}. Substituting eq.~\eqref{eq:U} into eq.~\eqref{eq:gradgen}, then carrying out a partial integration and introducing the radial distribution functions $g^{(ij)}(r;\boldsymbol{\lambda})$ and $g_0^{(ij)}(r)$ of the simulated and target ensembles, respectively, to take the required averages\cite{hansen:2006} gives
\begin{align}
    \frac{\partial S_{\rm rel}}{\partial \boldsymbol{\lambda}} &= \sum_{i=1}^n \sum_{j=1}^n 2 \pi \beta \frac{N_i N_j}{V} &\nonumber \\
    & \times \int_0^\infty \d{r} \, r^2 \left[g_0^{(ij)}(r) - g^{(ij)}(r; \boldsymbol{\lambda}) \right] \frac{\partial u^{(ij)}}{\partial \boldsymbol{\lambda}}.&
    \label{eq:gradg}
\end{align}
Both $g_0^{(ij)}$ and $g^{(ij)}$ can be calculated from particle trajectories, with the latter needing to be recomputed as $\boldsymbol{\lambda}$ is varied. We emphasize that eq.~\eqref{eq:gradg} is only valid for isotropic, pairwise interactions and when both the target and simulation ensembles are canonical. A similar derivation can be carried out for other ensembles\cite{chaimovich:jcp:2011, shell:jcp:2008, pretti:jcp:2021, lindquist:jcp:2019}. Anisotropic\cite{lieu:jcp:2022, lu:jctc:2014}, non-pairwise\cite{sanyal:jcp:2016}, or bonded interactions \cite{carmichael:jpcb:2012} would require additional distribution functions or direct evaluation of the ensemble averages in eq.~\eqref{eq:gradgen}. These calculations are all possible within the framework of our software but are not included in this work. We will discuss how the simulations needed to evaluate $g^{(ij)}$ are enabled by relentless in Section \ref{sec:design:simulate}.

The ability to readily evaluate eq.~\eqref{eq:gradgen} or eq.~\eqref{eq:gradg} with simulations naturally leads to application of local, gradient-based optimization methods to minimize $S_{\rm rel}$\cite{chaimovich:jcp:2011}. For example, the gradient descent method,
\begin{equation} \label{eq:steepestdescent}
    \boldsymbol{\lambda} \gets \boldsymbol{\lambda} - \alpha \frac{\partial S_{\rm rel}}{\partial \boldsymbol{\lambda}}, 
\end{equation}
is commonly used to iteratively update $\boldsymbol{\lambda}$ until satisfactory convergence is achieved, with $\alpha$ being the step-size hyperparameter. Variations of this method, as well as other gradient-based schemes, can also be employed. Some of us have also recently explored use of surrogate-based optimization to address computational challenges with gradient-descent minimization of the relative entropy \cite{petix:jctc:2024}. We will discuss optimization strategies further in Section \ref{sec:design:optimize}.

\section{Design and implementation}
\label{sec:design}
To enable optimizations that require running simulations, such as those outlined in Section~\ref{sec:REM}, relentless is organized into three modules: model, simulate, and optimize (Fig.~\ref{fig:overview}). The software uses an object-oriented design to ensure it is extensible to new applications and methods. We will now outline the main features we have developed in each of these modules that help perform more transparent, reproducible MD simulations for optimization.

\subsection{Model}
\label{sec:design:model}
The model module implements fundamental concepts related to defining a statistical mechanical ensemble, a potential energy function, and relationships between model parameters. The ensemble is a container for information about the thermodynamic state. Specifically, the ensemble includes the number of particles $N_i$ of each type $i$, the temperature $T$, the pressure $P$, and the volume $V$. The volume is defined as a spatial extent in two or three dimensions (with some abuse of language, as this extent is an area in two dimensions). In three dimensions, the most general extent that is supported is a parallelepiped specified according to either the standard HOOMD-blue \cite{anderson:compmatsci:2020} or LAMMPS \cite{thompson:compphyscom:2022} conventions, but specializations such as a cube are included for convenience. The equivalent parallelogram and square extents are implemented in two dimensions. The ensemble can additionally store the radial distribution function $g^{(ij)}$ between types $i$ and $j$. In practice, the ensemble is used both to help initialize simulations and to record results.

Particle interactions are currently modeled in relentless using short-ranged pairwise potentials. For inverse design, it may be necessary to combine one or more pairwise potentials to form the total interaction between a pair of particles \cite{lindquist:softmatter:2016}. To simplify the processes of combining pair potentials and implementing new pair potentials, we define a pair potential $u(r; \boldsymbol{\lambda}, r_{\rm min}, r_{\rm max})$ having a set of parameters $\boldsymbol{\lambda}$ that acts only when $r_{\rm min} \le r \le r_{\rm max}$ in terms of a nominal pair potential $u_0(r; \boldsymbol{\lambda})$ that is valid for all $r$. For notational convenience, we do not denote here that these potentials and parameters should all be specified between particles of every pair of types $i$ and $j$, but relentless supports this pairwise parametrization. We then restrict $u_0$ in a piecewise-continuous fashion
\begin{equation}
u(r) =
    \begin{cases}
    u_0(r_{\rm min}),& r < r_{\rm min} \\
    u_0(r),& r_{\rm min} \le r \le r_{\rm max} \\
    u_0(r_{\rm max}),& r > r_{\rm max}
    \end{cases},
    \label{eq:upiece}
\end{equation}
where we have explicitly shown only the dependence on $r$. The corresponding force $f$, which is needed to perform an MD simulation, is
\begin{equation}
f(r) = -\frac{\partial u(r)}{\partial r} =
    \begin{cases}
    0,& r < r_{\rm min} \\
    f_0(r),& r_{\rm min} \le r \le r_{\rm max} \\
    0,& r > r_{\rm max}
    \end{cases},
\end{equation}
where $f_0 = -\partial u_0/\partial r$ is the nominal force given by $u_0$. The partial derivative with respect to a parameter $\lambda_i$ is
\begin{equation}
\frac{\partial u(r)}{\partial \lambda_i} =
    \begin{cases}
    \dfrac{\partial u_0(r_{\rm min})}{\partial \lambda_i},& r < r_{\rm min} \\
    \dfrac{\partial u_0(r)}{\partial \lambda_i},& r_{\rm min} \le r \le r_{\rm max} \\
    \dfrac{\partial u_0(r_{\rm max})}{\partial \lambda_i},& r > r_{\rm max}
    \end{cases}.
    \label{eq:dudlambda}
\end{equation}
Inclusion of $r_{\rm min}$ and $r_{\rm max}$ as parameters of $u$ also requires their derivatives. The partial derivative of $u$ with respect to $r_{\rm min}$ is
\begin{equation}
\frac{\partial u(r)}{\partial r_{\rm min}} = \begin{cases}
-f_0(r_{\rm min}),& r < r_{\rm min} \\
0,& \textrm{otherwise}
\end{cases}
\label{eq:dudrmin}
\end{equation}
and with respect to $r_{\rm max}$ is
\begin{equation}
\frac{\partial u(r)}{\partial r_{\rm max}} = \begin{cases}
-f_0(r_{\rm max}),& r > r_{\rm max} \\
0,& \textrm{otherwise}
\end{cases}.
\label{eq:dudrmax}
\end{equation}
Note that $r_{\rm max}$ plays the role of the truncation distance on the force typical in MD simulations \cite{allen:2017}, so relentless also allows $u$ to be shifted to 0 at $r_{\rm max}$ by subtracting $u(r_{\rm max})$. This shifting does not affect $f$, but it adds $f_0(r_{\rm max})$ to the values of the parameter derivatives given in eqs.~\eqref{eq:dudlambda}--\eqref{eq:dudrmax}.

Our equations for the parameter derivatives are for the case where $u$ depends directly on $\boldsymbol{\lambda}$; however, it is possible for potentials to depend indirectly on the parameters used for design. For example, the Lennard-Jones pair potential,
\begin{equation}
u_{\rm LJ}(r) = 4\varepsilon\left[\left(\frac{\sigma}{r}\right)^{12} - \left(\frac{\sigma}{r} \right)^6 \right],
\end{equation}
has two direct parameters: the interaction energy $\varepsilon$ and the particle size $\sigma$. It is common practice to enforce an $r_{\rm max}$ on $u_{\rm LJ}$ for computational efficiency, with a typical choice of $r_{\rm max} = 3\,\sigma$ based on its functional form. Defining $u$ according to eq.~\eqref{eq:upiece} with $u_0 = u_{\rm LJ}$, the partial derivative of $u$ with respect to $\sigma$ must be evaluated using the total derivative
\begin{equation}
\left(\frac{\partial u}{\partial \sigma}\right)_{r; \varepsilon} = \left(\frac{\partial u}{\partial \sigma}\right)_{r; \varepsilon, r_{\rm max}} + \left(\frac{\partial u}{\partial r_{\rm max}}\right)_{r; \varepsilon, \sigma} \frac{{\rm d} r_{\rm max}}{{\rm d} \sigma}.
\end{equation}
In general, parameters can have nontrivial dependencies on other parameters, and these relationships must be accounted for using a systematic derivative propagation framework. relentless contains a framework for specifying independent variables, which can be manipulated directly, and dependent variables, which depend on one or more independent or dependent variables and can be automatically constructed from mathematical operations on other variables. The dependencies between variables are tracked in a directed acyclic graph that is used to evaluate the variables and perform automatic differentiation of one variable with respect to another \cite{margossian:datamining:2019}.

To implement these mathematical considerations in our software, we have defined a pair potential $u$ as an abstract base class in relentless. The base class has abstract energy, force, and parameter-derivative methods that must be implemented for specific functional forms of $u_0$. Only expressions for the derivatives of $u_0$ with respect to its direct parameters are required.  The base pair potential class uses automatic differentiation to evaluate the actual pair potential $u$, force $f$, and partial derivatives with respect to any variable. In our initial release, we have supported the Lennard-Jones potential, Yukawa (screened electrostatic) potential, Asakura--Oosawa (depletion) attraction, and an Akima spline potential. Multiple pair potentials can be defined in this way and combined when a simulation is run, as we will discuss next.

\subsection{Simulate}
\label{sec:design:simulate}
The simulate module is responsible for configuring and running a simulation using a model (Section \ref{sec:design:model}). The software design aims for a reusable, extensible interface that enable workflows compatible with different MD software packages. To achieve this, we have categorized the typical stages of a simulation workflow into three prototypical operations: initialization, simulation, and analysis. An initialization operation is responsible for basic configuration, such as establishing the initial particle configuration and setting up the particle interactions. The simulation operations execute one or more tasks, such as energy minimization or time integration of particle coordinates. Analysis operations, such as ensemble averaging, compute or store information from the simulation and are associated with a simulation operation. A complete list of simulation operations is available in the relentless documentation, but some representative examples include: initialization in a quasi-random configuration or from file; time integration using molecular (Newtonian) dynamics, Langevin dynamics, or Brownian dynamics; and calculation of ensemble-averaged thermodynamic properties or writing particle trajectories for visualization. Notably, we use the ensemble-averaging analysis operation to evaluate eq.~\eqref{eq:gradg}.

For each operation, we have defined a generic Python interface that is agnostic to a user's choice of simulation package. The user configures a sequence of operations, then the operations are translated into the package-specific commands and inputs needed to execute the operation at run time. A run is performed with the current values of the model parameters and with the user-configured potentials. Multiple potentials can be combined together; relentless accumulates the potentials into a tabulated format compatible with different simulation packages. This software design makes workflows easily reusable because the same operations can be performed with a different simulation package by making a one-line edit. relentless currently supports multiple versions of HOOMD-blue and LAMMPS as simulation packages.

To ensure the full performance of these simulation packages can be leveraged, relentless is compatible with distributed parallelism using the Message Passing Interface (MPI). relentless itself can be safely run with MPI if the simulation package can be initialized in parallel within Python, or it can externally run a parallelized simulation with an appropriate MPI-launched executable. Use of graphics processing units (GPUs) to accelerate simulations is also straightforward.

By choice, not all simulation operations are required to be supported by all simulation packages. We adopted this design to make it easier to extend relentless to new simulation packages, to support different versions of a given package, and to add new simulation operations that may be highly tailored to a specific workflow or package. Users requiring such customization can still take advantage of many of the established operations.

\subsection{Optimize}
\label{sec:design:optimize}
The optimize module defines and solves optimization problems.  Optimization requires an objective function to minimize, an algorithm to update model parameters, and convergence criteria to terminate the calculation. relentless currently supports $S_{\rm rel}$ as an objective function [eq.~\eqref{eq:srel}], but it is extensible to other user-defined objective functions. Calculation of objective functions can involve any sequence of steps, including running simulations (Section~\ref{sec:design:simulate}). For example, each evaluation of the gradient of $S_{\rm rel}$ [eq.~\eqref{eq:gradg}] launches an MD simulation to sample $g^{(ij)}$ in the simulation ensemble.

Given the focus of this work on relative entropy minimization, we have currently implemented basic gradient-based optimization strategies because only the gradient (and not the value) of $S_{\rm rel}$ is accessible to us. Gradient-based methods sometimes perform better when the gradient has components of comparable magnitude, so relentless performs its gradient-based optimization using scaled variables $x_i = \lambda_i/\Lambda_i$, where $\Lambda_i$ is the scale factor for $\lambda_i$. Applying this transformation during optimization allows users to employ simpler unscaled variables in the rest of their workflow. The partial derivative of an objective function $f$ with respect to $x_i$ is
\begin{equation}
\frac{\partial f}{\partial x_i} = \Lambda_i \frac{\partial f}{\partial \lambda_i}
\end{equation}
and the direction of descent $\hat{\mathbf{d}}$ is opposite the direction of the gradient with respect to the scaled variables $\mathbf{x}$,
\begin{equation}
\hat{\mathbf{d}} = -\frac{\partial f/\partial \mathbf{x}}{|\partial f/\partial \mathbf{x}|}.
\end{equation}
The gradient descent algorithm iteratively updates the scaled variables $\mathbf{x}_k$ at iteration $k$ to those at iteration $k+1$ by
\begin{equation}
\mathbf{x}_{k+1} = \mathbf{x}_k + d \hat{\mathbf{d}}_k
\end{equation}
where $d$ is a step size parameter. During the optimization, relentless respects any box constraints that have been placed on the values of the model parameters, and it performs iterative optimization until user-configured stopping criteria are satisfied.

Choosing $d = \alpha |\partial f/\partial \mathbf{x}|_k$ gives the standard steepest descent algorithm of eq.~\eqref{eq:steepestdescent}, where the size of each update to $\mathbf{x}$ is proportional to the magnitude of the gradient. Alternatively, choosing a constant $d$ gives a fixed-step-length update throughout the optimization. Steepest descent is numerically stable for sufficiently small $\alpha$, but slows down as it approaches a local minimum. Fixed-step descent avoids this slowdown but can have trouble navigating non-smooth functions that have a sharp local minimum with respect to one variable or achieving convergence when close to the minimum \cite{meza:compstat:2010}. In these cases, fixed-step descent tends to oscillate without making much progress. To address this issue, relentless can optionally use a line search during gradient descent. Using the user-provided $d$ as the maximum step size, an optimal $d_k$ for iteration $k$ is searched for on $0 < d_k \leq d$ such that
\begin{equation} \label{eq:curvecond}
\Delta(d_k) = -d \hat{\mathbf{d}}_k \cdot \left.\frac{\partial f}{\partial \mathbf{x}}\right|_{k+1}
\end{equation}
is sufficiently reduced below a user-specified tolerance, with the derivative being evaluated at the $\mathbf{x}_{k+1}$ corresponding to $d_k$. Because $\hat{\mathbf{d}}_k$ is a descent direction, $\Delta(0) > 0$, while $\Delta(d_k) < 0$ if $f$ is no longer decreasing in the direction of $\hat{\mathbf{d}}_k$ with the chosen $d_k$. Hence, choosing a step size such that $\Delta(d_k) \approx 0$ tends to decrease $f$ as much as possible in the direction $\hat{\mathbf{d}}_k$ before a new direction should be chosen. In practice, we accept the initial step $d$ if $\Delta(d) \ge c \Delta(0)$, where $c$ is a small constant; otherwise, we attempt to solve for $|\Delta(d_k)| \le c \Delta(0)$ using linear interpolation and bisection. This so-called curvature condition can be augmented with a criterion based on the value of $f$ but this is not suitable for minimization of $S_{\rm rel}$, which cannot be readily evaluated.  

In principle, any objective function and optimization algorithm can be implemented in relentless's extensible optimize module. In addition to $S_{\rm rel}$, objective functions based on free energy landscapes\cite{long:molsystdeseng:2018}, structural order parameters\cite{kumar:jchemphys:2019}, and scattering patterns \cite{coli:sciadv:2022} work well for optimizing particle interactions and often require simulations to evaluate. Optimizations focused on material properties, rather than microstructure, can also be performed with relentless by writing an objective function that uses simulations to measure, for example, mechanical response \cite{miskin:natmater:2013}, ionic conductivity \cite{kadulkar:jphyschemb:2021}, or light-matter interactions \cite{molesky:natphotonics:2018}. For some of these applications, gradient-free optimization methods may be better suited than gradient-based methods, and, in general, iterative gradient-based methods may have difficulty locating local or global minima for certain problems. In these cases, relentless could be integrated with more sophisticated optimization tools. For example, some of us have shown that surrogate models\cite{bhosekar:compcheme:2018, alizadeh:rengdes:2020} can greatly simplify relative entropy minimization \cite{petix:jctc:2024}, and relentless was used to collect the data needed to train these surrogate models. This strategy also suggests a route to interface relentless with open-source optimization software, such as Pyomo \cite{hart:mathprogcomp:2011, bynum:2021}, in the future.

\section{Examples}
\label{sec:examples}
We will now demonstrate the use of relentless to design pairwise interactions that produce a target structure in two example systems: (1) a binary fluid of nearly hard spheres and (2) a cubic diamond superlattice. These examples were selected because they highlight many features of the software design described in Section~\ref{sec:design}. Our analysis of these examples also provides some insight into numerical protocols for this inverse design problem.

\subsection{Binary hard-sphere fluid}
We first used relentless to determine the unknown diameters characterizing the interactions in a binary mixture of nearly hard spheres. The particle interactions were modeled using the repulsive Weeks--Chandler--Andersen (WCA) pair potential \cite{weeks:jcp:1971}
\begin{equation} \label{eq:WCApotential}
\beta u^{(ij)}(r)= \begin{cases}
\displaystyle   
4 \left[\left(\frac{\sigma_{ij}}{r}\right)^{12}-\left(\frac{\sigma_{ij}}{r}\right)^6 \right] + 1,& r \leq 2^{1/6} \sigma_{ij} \\
0,& {\rm otherwise}
\end{cases},
\end{equation}
where $\sigma_{ij}$ is the effective diameter for interactions between particles of types $i$ and $j$. For a binary mixture, there are 3 parameters: the self-interaction parameters $\sigma_{11}$ and $\sigma_{22}$, which are proportional to the diameters of particles of types 1 and 2, respectively, and the cross-interaction parameter $\sigma_{12}$, which is proportional to the mean of the two particle diameters. For our demonstration, we generated a target ensemble by specifying $\sigma_{11}$ as the unit of length and choosing $\sigma_{22} = 2.0\,\sigma_{11}$ and $\sigma_{12} = 1.5\,\sigma_{11}$. We then ran a Langevin dynamics simulation in LAMMPS (23 June 2022 version) using eq.~\eqref{eq:WCApotential} to compute the target radial distribution functions $g_0^{(11)}$, $g_0^{(12)}$, and $g_0^{(22)}$. We chose volume fractions $\phi_1 = 0.1$ and $\phi_2 = 0.2$ for types 1 and 2, respectively, in a cubic box width edge length $L = 40\,\sigma_{11}$ [$\phi_i = N_i \pi \sigma_{ii}^3/(6L^3)$], and we confirmed that a fluid phase was obtained under these conditions\cite{dijkstra:physrev:1999}. The simulation timestep was $0.005\,\tau$ and the friction coefficient was $0.1\,m/\tau$, where  $m$ is the particle mass and $\tau = \sqrt{m k_{\rm B} T/\sigma_{11}^2}$ is the unit of time. The particles were randomly initialized in the simulation box without hard-sphere overlap based on the diameters $\sigma_{11}$ and $\sigma_{22}$ and equilibrated for $10^3\,\tau$, then the radial distribution functions were computed up to separation distance $6\,\sigma_{11}$ with a bin spacing of $0.05\,\sigma_{11}$ from configurations sampled every $10^2\,\tau$ during a $10^5\,\tau$ production simulation.

\begin{figure}
    \includegraphics{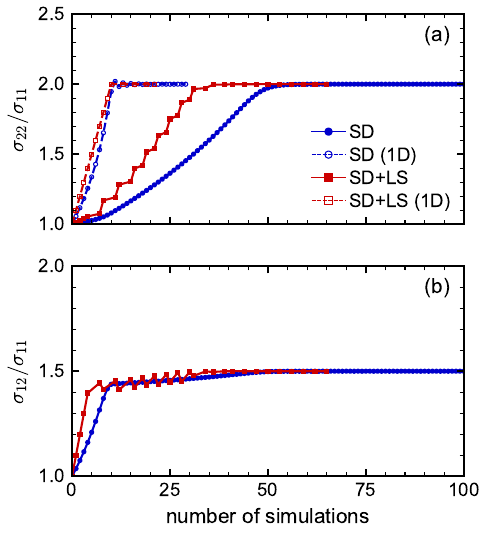}
    \caption{Binary hard-sphere fluid optimization using either a standard steepest descent (SD, blue circles) or fixed-step steepest descent with line search (SD+LS, red squares) approach. The evolution of independent variables (a) $\sigma_{22}$ and (b) $\sigma_{12}$ is shown as a function of number of simulations performed. The solid lines with filled symbols are designs that considered both $\sigma_{22}$ and $\sigma_{12}$ as independent variables, while the dashed lines with open symbols are designs that considered only $\sigma_{22}$ as an independent variable.}
    \label{fig:hs_params}
\end{figure}

We then performed relative entropy minimization to design $\sigma_{12}$ and $\sigma_{22}$ starting from an naive initial guess $\sigma_{12} = \sigma_{22} = 1.0\,\sigma_{11}$. The simulation protocol was similar to that used to generate the target, except that an energy minimization was performed before equilibration to an energy tolerance of $10^{-4}\,k_{\rm B}T$ and a force tolerance of $10^{-6}\,k_{\rm B}T/\sigma_{11}$. We minimized the intensive relative entropy, $s_{\rm rel} = S_{\rm rel}/V$ to a tolerance of $10^{-4}$ for each component of the gradient. Figure \ref{fig:hs_params} shows the evolution of these parameters using two optimization approaches: a standard steepest descent (SD) [eq. \eqref{eq:steepestdescent}] with $\alpha = 0.2$ and a fixed-step steepest descent supplemented by a line search (SD+LS) with $d = 0.1$, $c = 0.1$, and a maximum of 3 iterations for the line search. We compared the two methods in terms of number of simulations required for convergence. Both approaches successfully converged to the known target parameter values, but the SD+LS approach required significantly fewer simulations than the SD approach because it could sometimes take large step sizes between iterations.

\begin{figure}
    \includegraphics{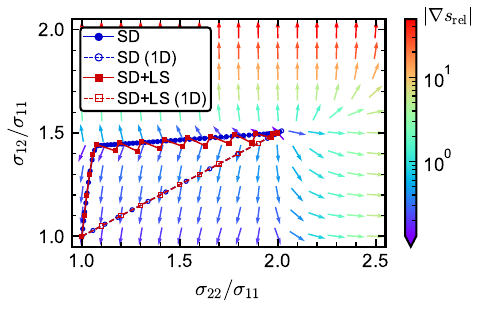}
    \caption{Evolution of $\sigma_{22}$ and $\sigma_{12}$ imposed on the relative entropy gradient field for the same designs as in Fig.~\ref{fig:hs_params}.}
    \label{fig:hs_grad}
\end{figure}

Although $\sigma_{12}$ can be adjusted independently of $\sigma_{22}$, it may be helpful to use physical intuition to constrain it to the arithmetic mean of the particle diameters, i.e., $\sigma_{12} = (\sigma_{11} + \sigma_{22})/2$, in order to eliminate an independent variable from the optimization and reduce it to a simpler one-dimensional (1D) problem. We performed an additional relative entropy minimization using this mixing rule to constrain $\sigma_{12}$; for this optimization, we were able to use a larger $\alpha = 0.5$ for the SD approach. We found that significantly fewer simulations were needed to converge the design [Fig.~\ref{fig:hs_params}(a)], with the SD+LS approach requiring slightly fewer simulations than the standard SD. To better understand this faster convergence, we characterized the gradient field of the relative entropy and plotted the evolution of $\sigma_{12}$ and $\sigma_{22}$ both without and with the constraint on $\sigma_{12}$ (Fig.~\ref{fig:hs_grad}). The optimization of $\sigma_{12}$ and $\sigma_{22}$ took an indirect path to the minimum, first quickly increasing $\sigma_{12}$ into a ``valley'' before increasing $\sigma_{22}$ more slowly. In contrast, the 1D optimization of only $\sigma_{22}$ could take a more direct route that did not follow the two-dimensional gradient field because $\sigma_{12}$ was constrained. 

Although this example is a contrived demonstration because the target parameters were already known, it highlights several important capabilities of relentless. First, both the cutoff distance and mixing rule in eq.~\eqref{eq:WCApotential} required use of the chain rule to propagate derivatives with respect to $\sigma_{22}$ to evaluate eq.~\eqref{eq:gradg}. The parameters that were fed to the simulation also needed to be continually updated as the independent variables were adjusted. Last, we were able to easily switch between optimization approaches and our choice of independent variables with minimal changes to our code. All of these possible complications were handled automatically by relentless.

\subsection{Cubic diamond superlattice}
\begin{figure*}
    \includegraphics{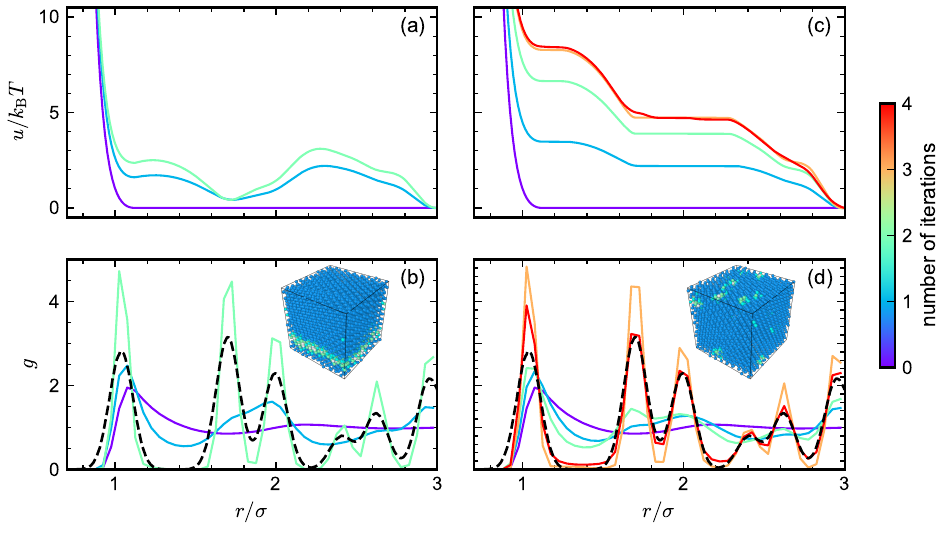}
    \caption{Cubic diamond optimization using a WCA steric repulsion plus either (a)--(b) an unconstrained difference spline (DS) potential or (c)--(d) a DS potential constrained to be purely repulsive. (a) and (c) show the evolution of the total pair potential $u$, while (b) and (d) show the evolution of the radial distribution function $g$ as the optimization progressed. The target structure is shown as a dashed line in (b) and (d). The inset snapshots are from the final iteration with particles colored according to their local morphology: cubic diamond (dark blue), nearest neighbor of cubic diamond (light blue), second-nearest neighbor of cubic diamond (green), and disordered (white)\cite{stukowski:modsimmatscieng:2009, maras:compphyscom:2016}.}
    \label{fig:dmnd_evo}
\end{figure*}
We then applied relentless to design a pair potential for colloidal particles that self-assemble a cubic diamond superlattice. Cubic diamond is an open lattice with a closest-packed volume fraction of only $\phi \approx 0.34$ and is desirable for photonics applications because cubic diamond lattices made from colloidal particles can have a complete photonic band gap in the visible range \cite{yablonovitch:joptsocamb:1993, moroz:physrevlett:1999}. However, cubic diamond is not typically a stable crystalline phase for spherical particles interacting through short-ranged attractive and/or repulsive forces, which usually prefer to form denser lattices \cite{schultz:jcp:2018, ahmed:physrev:2009, kaldasch:langmuir:1996}. Hence, finding an isotropic pair potential that self-assembles cubic diamond is a non-trivial design problem.

To generate the target ensemble, we tethered 8000 particles with nominal diameter $\sigma$ to sites of a cubic diamond lattice at a nominal volume fraction of 0.30, which is below close-packing. The tethers were harmonic potentials,
\begin{equation} \label{eq:tether}
\beta u_{\rm t}(\mathbf{r}) = \frac{k_{\rm t}}{2} \left| \mathbf{r} - \mathbf{r}_{0} \right|^2 ,
\end{equation}
where $\mathbf{r}$ is the position of the particle attached to the lattice site at $\mathbf{r}_{0}$ and $k_{\rm t} = 25\,\sigma^{-2}$ is the spring constant for the tether. The target $g_0$ is then straightforward to generate using either Gaussian-distributed random displacements or a short simulation. We ran Brownian dynamics in HOOMD-blue (version 2.9.7) with timestep $0.001 \tau_{\rm D}$, where $\tau_{\rm D} = \beta \gamma \sigma^2/4$ is a diffusive time scale and $\gamma$ is the drag coefficient. We initialized particles on cubic diamond lattice sites, equilibrated for $10 \tau_{\rm D}$, sampled configurations every $1\,\tau_{\rm D}$ over $100\,\tau_{\rm D}$, and computed $g_0$ up to a distance of $5\,\sigma$ with a bin spacing of $0.005\,\sigma$.

For the design, we represented the pair potential $u$ between particles as a combination of a fixed WCA core [eq.~\eqref{eq:WCApotential} with diameter $\sigma$] representing steric repulsion and an interpolating Akima spline with a set of knot points $\{(r_i, u_i)\}$ that could be designed. We equally spaced 40 knots in the interval $2^{1/6}\,\sigma \le r \le 3\,\sigma$. Following guidance from prior work, we defined the independent variables that we optimized using a ``difference spline'' (DS) approach, where the independent variables were the differences $\Delta_i = u_i - u_{i+1}$ between consecutive knot values with the final knot value fixed at zero. Langevin dynamics simulations were performed using timestep $0.005\,\tau$ and friction coefficient $0.1\,m/\tau$. ($\tau$ now uses $\sigma$ as the unit of length rather than $\sigma_{11}$.) The simulation protocol was: initialize the particles randomly in the box without hard-sphere overlap based on the diameter $\sigma$, equilibrate at an elevated temperature $T_{\rm eq} = 2.5\,T$ for $5 \times 10^3\,\tau$, linearly ramp the temperature from $T_{\rm eq}$ to $T$ over $2.5 \times 10^4\,\tau$, then compute the radial distribution function up to separation distance $3\,\sigma$ with a bin spacing of $0.05\,\sigma$ from configurations sampled every $50\,\tau$ during a $5 \times 10^3\,\tau$ production simulation. The minimization of $s_{\rm rel}$ was performed using the SD+LS approach with $\alpha = 1.0$, $c = 0.1$, and a maximum of 3 iterations for the line search to a loose tolerance of 0.5 for each component of the gradient.

Using this approach, we were able to quickly design a potential that self-assembled the cubic diamond lattice [Figs.~\ref{fig:dmnd_evo}(a) and \ref{fig:dmnd_evo}(b)]. Peaks in the target $g_0$ corresponded to attraction in the designed $u$, while minima in the target $g_0$ tended to correspond to repulsion in the designed $u$. However, prior work has shown that it is possible to self-assemble crystalline lattices using purely repulsive potentials \cite{lindquist:jcp:2016, jadrich:jcp:2017}, so we forced the DS to be purely repulsive by constraining the differences between knot values to be nonnegative. We were able to design a purely repulsive potential [Figs.~\ref{fig:dmnd_evo}(c) and \ref{fig:dmnd_evo}(d)] that also self-assembled cubic diamond, finding that attractive minima in $u$ tended to become flat, which is consistent with previous studies \cite{lindquist:jcp:2016, jadrich:jcp:2017}.

\begin{figure}
    \includegraphics{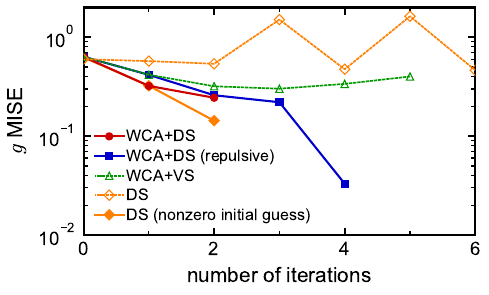}
    \caption{Mean integrated square error (MISE) in $g$ for different design approaches for cubic diamond. WCA+DS is the WCA steric repulsion with an unconstrained difference spline (DS). WCA+DS (repulsive) is the same as WCA+DS, but the DS is constrained to be repulsive. WCA+VS is the WCA steric repulsion with an unconstrained value spline (VS). DS is only the difference spline initialized as either zeros or the repulsive potential of ref.~\citenum{lindquist:jcp:2016}. The solid lines with filled symbols are designs that converged, while the dashed lines with open symbols are designs that did not converge.}
    \label{fig:dmnd_err}
\end{figure}

After performing these designs, we then considered how variations on our approach might affect success and/or rate of convergence. To be able to compare different approaches, we computed the mean integrated square error (MISE) between the simulated $g$ and target $g_0$,
\begin{equation}
g\:\mathrm{MISE} = \frac{3}{R^3} \int_0^{R} \!\!\! {\rm d}r \, r^2 \left[ g(r) - g_0(r) \right]^2,
\end{equation}
where $R = 3\,\sigma$ was the truncation distance for both $u$ and our calculation of $g$. In agreement with Fig.~\ref{fig:dmnd_evo}, the MISE decreased as the optimization progressed, and the final MISE was smaller for the potential that was constrained to be purely repulsive than for the unconstrained one (Fig.~\ref{fig:dmnd_err}).

We then compared our DS approach to a naive ``value spline'' (VS) approach that used the values of the spline knots directly as independent variables. For an equivalent potential, the VS and DS representations give different gradients of $S_{\rm rel}$ with respect to $\boldsymbol{\lambda}$ and so may converge differently. Indeed, the optimization of the VS did not converge within the allowed number of iterations and the MISE even increased as the optimization progressed. Other studies have noted that the DS approach performs better than the VS approach for relative-entropy optimization of spline potentials \cite{lindquist:jcp:2016, jadrich:jcp:2017}. We speculate the better performance of the DS approach is because $\partial u/\partial \Delta_i$ leads to nonlocalized changes in the potential due to chain rule propagation of derivatives, but $\partial u/\partial u_i$ is nonzero and leads to changes in the value of the potential only locally around $r_i$.

\begin{figure}
    \includegraphics{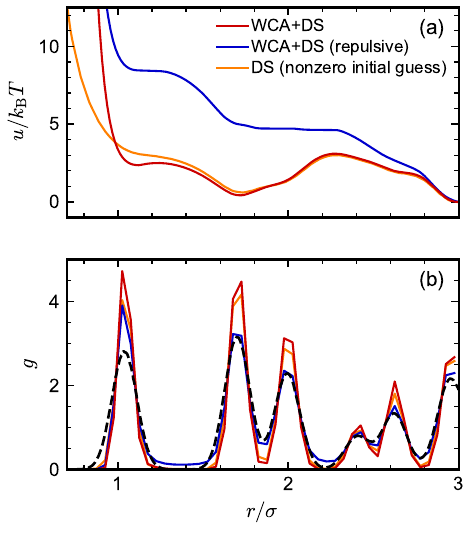}
    \caption{(a) Pair potential $u$ and (b) radial distribution function $g$ of the converged designs in Fig.~\ref{fig:dmnd_err}. The target structure is shown as a dashed line in (b).}
    \label{fig:dmnd_fin}
\end{figure}

We then considered the effect of including the WCA potential to account for steric repulsion instead of using a spline potential to represent the entire interaction. We extended the DS to $10^{-6}\,\sigma \le r \le 3\,\sigma$ and increased the number of knots to 60. We initialized the potential either to zero or to the repulsive form recommended in the supporting information of ref.~\citenum{lindquist:jcp:2016}. We found that the convergence of the design was highly sensitive to this initial guess. The potential that was initialized to zero did not converge within the allowed number of iterations, whereas the potential that was initialized with steric repulsion converged at a similar rate to the DS with WCA repulsion. The converged design had a softer repulsion at small $r$ compared to the WCA repulsion and was able to achieve a somewhat smaller MISE (Fig.~\ref{fig:dmnd_fin}), likely due to the additional flexibility in setting the repulsion.

This example highlights relentless's ability to combine and change the parametrization of pair potentials. Our numerical experiments using relentless show that difference splines seem to perform better than value splines and that steric repulsions that are known to be present should either be explicitly fixed in a pair potential or implicitly included as an initial guess of a spline potential that will be designed. We note that the self-assembled superlattices contained some defects [insets of Figs.~\ref{fig:dmnd_evo}(b) and \ref{fig:dmnd_evo}(d)]. We speculate that these defects may be a result of kinetic trapping in the simulation protocol that might be improved using temperature annealing or an alternative assembly strategy (e.g., slow volume compression).

\section{Discussion}
\label{sec:Discussion}
We have presented relentless, a Python package developed to easily incorporate molecular simulations into optimization workflows, such as those used for computational materials design. relentless has a modular, object-oriented software design that connects common tasks in molecular modeling, MD simulation, and basic optimization. Of particular note is relentless's standardized framework for specifying the operations in a simulation workflow to be executed natively using different packages, which simplifies the process of integrating MD simulation in larger workflows or porting work to different computational resources. We have also shown two examples to demonstrate relentless's capabilities for inverse design of pair potentials, including the ability to configure potential energy functions, describe complex parameter dependencies and constraints,  and select between different optimization approaches. By offering a standardized interface for both MD simulation and optimization, relentless has the potential to streamline the implementation of new inverse design algorithms that can ultimately help accelerate the discovery of novel materials with tailored properties and functionalities.

relentless's software design also promotes TRUE principles \cite{thompson:jmolphys:2020} for scientific software. Its open-source, standardized interface promotes sharing of transparent, reproducible workflows; for example, in creating our examples, we transferred simulation scripts between two high-performance computing resources with different computing hardware and MD software available, with only minimal changes required. relentless's object-oriented design also empowers users to both reuse standard elements and to make new ones to customize their workflow. relentless follows best-practices for scientific software development, including use of version control, continuous integration testing, and publicly hosted documentation. In future, we anticipate that relentless will be expanded to support additional simulation engines (e.g., GROMACS and OpenMM), additional potential energy functions (e.g., bonded interactions), and more objective functions and optimization strategies.

\section*{Supplementary Material}
See the supplementary material for relentless inputs for the examples in Section~\ref{sec:examples}.

\begin{acknowledgments}
ANS was supported by the Welch Foundation (Grant No.~F-1696). CLP was supported by the U.S. Department of Education, GAANN Program under Award No.~P200A210047 and a fellowship from The Molecular Sciences Software Institute under National Science Foundation Award No.~CHE-2136142. ZMS was supported by an Arnold O. Beckman Postdoctoral Fellowship. This work was completed with resources provided by the Texas Advanced Computing Center (TACC) at The University of Texas at Austin and the Auburn University Easley Cluster.
\end{acknowledgments}

\section*{Data Availability}
The source code for relentless is publicly available at https://github.com/mphowardlab/relentless, and relentless is publicly distributed through PyPI and conda-forge. The other data that support the findings of this study are available from the authors upon reasonable request.

\bibliography{references}

\end{document}